\documentclass[fleqn,usenatbib]{mnras}

\DeclareRobustCommand{\VAN}[3]{#2}
\let\VANthebibliography\thebibliography
\def\thebibliography{\DeclareRobustCommand{\VAN}[3]{##3}\VANthebibliography}

\usepackage{graphicx}	
\usepackage{amsfonts}
\usepackage{amsmath}	
\usepackage{amssymb}	
\usepackage{lscape}
\usepackage{newtxtext,newtxmath}
\usepackage{natbib}

\title[]{Clues from  4U 0142+61 on supernova fallback disc formation and precession}

\author[C. Grimani]{
Catia Grimani,$^{1,2}$\thanks{E-mail: catia.grimani@uniurb.it}
\\
$^{1}$DiSPeA, University of Urbino Carlo Bo, Urbino (PU), Italy\\
$^{2}$Italian Institute for Nuclear Physics, Florence, Italy\\
}

\date{Accepted XXX. Received YYY; in original form ZZZ}

\pubyear{2015}

\begin{document}
\label{firstpage}
\pagerange{\pageref{firstpage}--\pageref{lastpage}}
\maketitle

\begin{abstract}
The NuSTAR experiment detected a  hard X-ray emission (10-70 keV) with a period of 8.68917 s and a pulse-phase modulation at 55 ks, or half this value, from the anomalous X-ray pulsar (AXP) 4U 0142+61. It is shown here that this evidence  is naturally explained by the precession of a Keplerian supernova fallback disc  surrounding  this AXP. It is also found that the  precession of discs formed around young neutron stars  at distances larger than those considered in the past, may constitute almost neglected sources of  gravitational waves with frequencies belonging to the sensitivity  bands of the future space interferometers LISA, ALIA, DECIGO and BBO. In this work the gravitational wave emission from precessing fallback discs possibly formed around young pulsars such as Crab in a region extending beyond  8$\times$10$^{7}$ m  from the pulsar surface is estimated.  It is also evaluated the role that infrared radiation emission from circumpulsar 
discs may play in contributing to  Inverse Compton Scattering of  TeV energy positrons and electrons.
Extensive observational campaigns  of disc formation around young and middle aged pulsars  may also contribute to solve the long-standing problem of a pulsar origin for the  excess of positrons in cosmic rays observed near Earth above 7 GeV. In the near future the James Webb telescope, with unprecedented near and mid-infrared observation
capabilities,  may provide direct evidence of a large sample of supernova fallback discs.
\end{abstract}




\section{Introduction}

Monte Carlo simulations  indicate that supernova fallback discs may likely form near the light cylinder (LC)  of pulsars with birthperiods $>$ than 40 ms \citep{jiang}. Observational clues reported in \cite{gri16} indicate that this is the case for the majority of young pulsars.
However, depending on the angular momentum of fallback matter, discs may also form at different distances from the pulsar  \citep{currie}. 
The role  of  disc formation   in affecting the evolution of  pulsars \citep[][]{Benli_2017} and magnetars was extensively discussed  in the literature to explain observations from  soft gamma repeaters   \citep[SGRs,][]{tong}, anomalous X-ray pulsars   \citep[AXPs,][]{chatterjee,Alpar_2001,Ertan_2009}, rotating radio transients  
\cite[RRATs,][]{li06,Gencali_2020},  X-ray dim isolated neutron star \citep[XDINS,][]{Ertan_2014,xdins} and central compact objects \citep[CCOs,][]{erdeve,Benli_2018}. 

The  presence of a disc was claimed  around the  AXP 4U 0142+61 on the basis of the Spitzer satellite observations and earlier data \citep{wang,Ertan_2007}.  While this discovery was questioned in the literature \citep{durant}, 
the  NuSTAR experiment detected an X-ray emission modulated at 55 ks, or half this value, from the same  AXP  \citep{makishima}. It is  shown here that this modulation is compatible with  the presence of a Keplerian precessing disc with the characteristics  observed by  \cite{wang} hiding periodically the emission region. The possible formation of a disc around the magnetar 1E 2259+586  was also reported in \cite{kaplan}. 
A hundreds of AU region emitting infrared radiation   around the X-ray thermal isolated neutron stars (XTINS) RX J0806.4-4123 was  detected with the Hubble telescope  \citep{posselt}.
This observation is compatible with the  presence of a huge disc surrounding the pulsar.

The presence of a disc around the RRAT B0656+14 was  reported in \cite{perna}. In  \cite{gri13}  it was pointed out  that a precessing  disc could have explained the transient emission from this pulsar.

In the past, however, several campaigns of observations did not allow  to detect a large sample of  circumpulsar discs \citep{wolszczan}, possibly because of low instrument sensitivities  and  disc inclinations with respect to detectors field of view.

The TeV pulsed photon emission observed with the MAGIC experiment \citep{ansoldi}  in a region   extending beyond 8$\times$10$^7$ m  from the Crab pulsar  is consistent  with TeV e$^+$ and e$^-$ scattering infrared photons to TeV energies. In the case of  disc formation in that region both pulsar and disc infrared radiation may contribute to this process. The hypothesis of a disc surrounding  Crab  just beyond  the LC was  discussed  in  \cite{menou}. 

A step forward in the knowledge of pulsar and magnetar environments may also provide precious insights on positron observations in cosmic rays gathered near Earth showing an increasing excess of these particles with respect to the estimated secondary component
above 7 GeV, with a drop-off just below 300 GeV  and a possible cut-off at about 800 GeV \citep{amspos,caprioli,huang}. While an
 astrophysical or an exotic origin of the excess of positrons in cosmic rays remains  to be settled (see for instance \cite{harding,zhang,gri04,gri07,feng,manconi}),  observations  are compatible with  
sources of TeV e$^-$- e$^+$ located 
between 100 and 800 pc from the Solar System \citep{attallah}.  

The James Webb telescope  \citep[JWT,][]{gardner}, scheduled to launch in the  fall 2021, will definitely open the infrared observational window on  disc formation around magnetars and pulsars.

Following the NuSTAR  observations,  this paper  
  aims to evaluate the consequences  of   Keplerian discs precessing around pulsars and magnetars     
in  generating gravitational waves with frequencies belonging to the sensitivity bands of the future LISA \citep{lisa}, ALIA \citep{wtni}, BBO  \citep{cutler}, and DECIGO  \citep{decigo} space interferometer  sensitivity bands. 

In  \cite{gri16} the gravitational wave emission from  precessing discs formed near the LC of young and middle aged pulsars was considered and discussed. In the same paper it was also shown that  gravitational waves emitted by these precessing discs  could have been detected 
with the second generation LISA-like space interferometers such as BBO and DECIGO,   presenting maximum sensitivities near 1 Hz.  The NuSTAR  observations are consistent with the scenario of disc formation and precession proposed in \cite{gri16} but  may also suggest that  disc formation occur at larger distances than previously considered.  DECIGO remains the favored interferometer to detect gravitational waves  generated by precessing discs  formed  around pulsars possibly up to  distances exceeding tens of kpc from Earth. In particular, 
depending on the  actual dimensions and masses  of   discs and precession angles, observations carried out with the future space  interferometers may  help in composing the
 puzzle of disc role  in affecting the environment of neutron stars (NSs) especially in respect to high-energy electrons and positrons undergoing inverse Compton (IC) scattering  of soft photons before escaping the source region \citep{gri13,gri16}. 

In Section 2 characteristics of  the disc observed  around 4U 0142+61 and  speculations about a disc possibly formed around Crab are reported. In Section 3 the  precession of a disc around 4U 0142+61  is considered to explain the modulated X-ray emission observed from this magnetar.  Alternative scenarios involving pulsar free and forced precession have been also considered. In Section 4  gravitational wave emission from precessing discs around pulsars and magnetars is re-estimated following the observations of the NuSTAR experiment. Finally, in Section 5 the  infrared emission from  a disc formed around a young pulsar such as Crab is compared to  that of the host neutron stars in the disc region in order to estimate if TeV electron and positron propagation  could be affected by the disc presence.   

\section{Fallback disc characteristics}

\subsection{4U 0142+61}
The disc surrounding the  AXP 4U 0142+61  observed by Wang, Chakrabarty and Kaplan \cite{wang} has  a mass of  5.97$\times$10$^{25}$ kg and 
presents  inner ($R_I$) and outer ($R_O$) radii of 2.02$\times$10$^{9}$ m
and 6.75$\times$10$^{9}$ m, respectively. The disc height (h) is set here equal to 0.035$\times$$R_I$ according to \cite{refh}.
 The  observed disc  temperature  ($T_D$) is  920 K. The associated blackbody  radiation according to  Wien's law, $\lambda$$_{max}$T=2.898$\times$10$^{-3}$ m K with $\lambda$$_{max}$=3.15$\times$$10^{-6}$ m, peaks in the mid-infrared range.
The AXP 4U 0142+61 
 has a surface temperature of 0.309$\pm$0.001 keV  equivalent to a blackbody temperature ($T_{NS}$) of 3.59$\times$10$^6$ K  \citep{guvert}. By using the Stefan-Boltzmann law, $\sigma$T$^4$= 5.67$\times$10$^{-8}$ T$^4$ W m$^{-2}$,   and by considering 10 km for the magnetar radius ($R_{NS}$) the  temperature of the disc ($T_D$) illuminated by the pulsar can be inferred from: 
 
\begin{equation}
 \frac{R_{NS}^2}{R_{I}^2}\  \sigma \  T_{NS}^4 A_{D} =  A_{D1}\ \sigma T_{D}^4, 
\end{equation}

\noindent where  $A_{D}$ and $A_{D1}$ are the area of the disc illuminated by the pulsar and the  area of the disc emitting as a blackbody, respectively. 
Several uncertainties limit the precision of this estimate:   a) the actual radius of  pulsars/magnetars ranging between 10 and 14 km for a typical 1.4 solar mass star \citep{steiner}, b) the exact pulsar/magnetar blackbody radiation 
at the equatorial region  (where the disc plausibly forms)  being  about 30\% smaller than that of the hot spots and polar 
cap  \citep{aguilera} and c) the actual thickness and energy absorbed by the disc \citep{beinli,shakura,lu}.
Finally, the NS blackbody photon gravitational redshift would be of the order of 10\%. Even by taking into account these considerations, 
the  conclusion  about the energy of photons reaching the disc would not change sensibly.
 The detected blackbody temperature of the disc would result of 920 K from equation (1) if, as plausible,  the energy  absorbed by the disc surrounding the magnetar is  5\% of the illuminating energy.
\subsection{Crab}

 The MAGIC experiment observed a  pulsed gamma-ray emission   from the Crab pulsar up to 1.5 TeV energy \citep{ansoldi}.  
The  gamma-ray emission pulse profile presents two peaks and  the energy differential fluxes show spectral indices close to 3.
These observations are consistent with electrons and positrons with Lorentz factors $\gamma$ $>$ 5$\times$10$^6$  undergoing IC scattering 
with low-energy photons, typically infrared radiation, mainly concentrated at distances well beyond 50 LC radii from the Crab surface.  The process that accelerates electrons and positrons is not well understood  \citep{ansoldi}, but the infrared emission from a disc, in addition to pulsar magnetospheric and/or pulsed infrared radiation, may play a role  in affecting the high-energy electron and positron propagation.  
 We are aware that this is only a possible scenario but it is proposed by considering that the formation of a disc around Crab should be favoured well beyond the typical LC distance as proposed by \cite{menou} because the temperature falls below 2000 K,  similarly to the case of 4U 0142+61. 

The  upper limit of the blackbody temperature  of the Crab pulsar is 2.1 MK \citep{tennant}. 
According to the Wien's law, the blackbody radiation  peaks in the X-ray band at 143
 eV  and the average energy  is:

\begin{equation}
\epsilon_{avg} =  \frac{\sigma c^2 h^3 T}{2.405\ 2 \pi \ k^3}= 490\ eV.
\end{equation}

As recalled above, the  gravitational redshift would not affect 
significantly  the described  scenario.
By using the same considerations applied to 4U 0142+61 in the previous Section, if a disc  surrounds  Crab   beyond 50  LC  distance from the pulsar surface, the  disc inner radius would be > 8$\times$$10^{7}$  m   and the thickness could be set to 2.8$\times$10$^6$ m as a lower limit. 
Given the blackbody temperature of Crab 
and a disc absorbing 10$^{-3}$ of the illuminating X-ray energy, the nominal inner temperature of the disc would be about 1800 K according to equation 1.
 The blackbody radiation of discs of 1800 K and 920 K is reported in Figure 1  for comparison with Figure 3 in \cite{wang}.

\begin{figure}
\begin{center}
\includegraphics[width=\columnwidth]{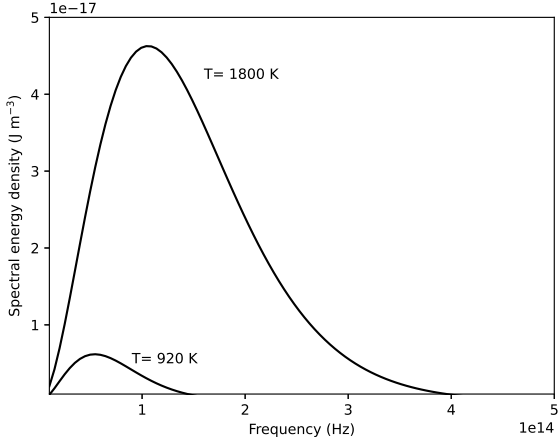}
\end{center}
\caption{\label{}  Blackbody radiation spectral energy density for disc temperature of 920 K and 1800 K.}
\label{figure1}
\end{figure} 
 
An absorption of the illuminating energy of 10$^{-2}$  as  indicated by the blackbody radiation of the disc surrounding 4U 0142+61 
would lead to a similar temperature of the disc in case the same  would be  thicker or larger than assumed. 
 Moreover, the disc may extend well beyond 50 LC where the temperature has  values similar or smaller than that  of the disc observed around  4U 0142+61. The comparison between disc and pulsar infrared emission in the disc region is discussed in Section 5.  

\section{Implications of a Keplearian precessing  disc surrounding  4U 0142+61}

The  modulated X-ray emission  at 55(27.5)  ks observed with the NuSTAR experiment from 4U 0142+61 was explained in \cite{makishima} with a magnetar free precession induced by the  star 
magnetic field of about 10$^{16}$ G. Nevertheless, this magnetic field value   exceeds by far  previous estimates reported in the literature of (4.75$\pm$0.02)$\times$10$^{14}$ G \citep{guvert}.   Forced precession of a pulsar due to the action of a surrounding accretion or  fallback disc  was proposed by  \cite{tong1}. If this process explains the NuSTAR observations from 4U 0142+61, the angular momentum of the disc must be $>$ than that of the pulsar and the precession period is expected to appear  consistent with the observations. By defining  $I_{4U\ 0142+61}$ and $I_{D}$  the moments of inertia of the magnetar and of the disc, respectively, as follows:

\begin{equation}
 I_{4U\ 0142+61}= \frac{2}{5} M_{4U\ 0142+61} R_{4U\ 0142+61}^2 = 1.12\times10^{38} kg\ m^2
\end{equation}

\begin{equation}
 I_D= \frac{1}{2} M_D (R_I^2+R_O^2) = 1.48\times10^{45} kg\ m^2,
\end{equation}

where $R_{4U\ 0142+61}$=$R_{NS}$ and
$M_{4U\ 0142+61}$=2.8$\times$10$^{30}$ kg, then the angular momenta of the magnetar ($L_{4U 0142+61}$) and of the disc ($L_D$) can be calculated as indicated below:

\begin{equation}
 L_{4U 0142+61}= I_{4U 0142+61} \omega_{4U 0142+61},
\end{equation}

\begin{equation}
  L_{D}= I_{D}\omega_{D}.
\end{equation}

The angular frequency of the magnetar is $\omega$$_{4U 0142+61}$=0.723 Hz and the Keplerian angular frequency of  the disc, $\Omega_K$=$\omega_{D}$ is determined
 from the equation  $V$=$\Omega_K$$R_I$, where $V$= $\sqrt{(GM_{4U 0142 +61})/R_I}$ is the disc velocity at the inner radius, $R_I$,
and 
$G$ is the universal gravitational constant equal to 6.67$\times$10$^{-11}$N m$^{2}$ 
kg$^{-2}$. As recalled in the previous Section,   the inner radius of the disc around  4U 0142+61 is 2.02$\times$10$^{9}$ m, therefore $V$ is 3.0$\times$10$^{5}$ m s$^{-1}$
 and $\Omega_K$= $\sqrt{(GM_{4U 0142+61})/R_I^3}$=1.505$\times$10$^{-4}$ s$^{-1}$. As a result,

\begin{equation}    
 L_{4U\ 0142+61}=8.10\times 10^{37} kg\ m^2 s^{-1}   
\end{equation}

and

\begin{equation}
  L_{D}= 2.23\times10^{41} kg\ m^2 s^{-1}.
\end{equation}


The disc may induce the magnetar under precession, however  the precession period,  estimated according to \cite{qiao} as:

\begin{equation}
 P= \frac{2 \pi}{|\dot \phi|},
\end{equation}
 
where

\begin{equation}
|\dot \phi|= \frac{45 M_D cos\theta' \omega_{4U 0142+61}}{32 \pi \rho R_I^3},
\end{equation}

is 2.77$\times$10$^{20}$ s, by reasonably assuming 
in the above equation  $\theta'$=15$^{\circ}$ and $\rho$=10$^{17}$ kg m$^{-3}$ for the neutron star density. It is recalled here that $\theta'$ represents the angle formed by  the rotation axis of the pulsar and the  perpendicular  to the disc plane. The result obtained is definitely not in agreement with the NuSTAR observations. The precession period increases with $\theta'\to \pi/2$ or if the disc radius increases.
The model by \cite{tong1}  applies better to scenarios involving  heavy discs formed near young pulsars.


In the  paper by  \cite{makishima} our previous work \citep{gri16} was mentioned to point out that our results were not in agreement with the NuSTAR observations. However, this is not the case.
  In the following it is shown that the free precession of a Keplerian disc (due to a misalignment  of the symmetry
 and the angular momentum axes) with the characteristics reported in Section 2  would  naturally explain the
 NuSTAR observations.
If $I_1$,  $I_2$  and $I_3$  are  the principal  moments of inertia of the disc with respect to
the principal axes,  $x_1$, $x_2$ and $x_3$ fixed in the disc and $\Omega_3$ is the angular velocity along the symmetry axis $x_3$, then
$\Omega_3$ is commonly assumed equal to the Keplerian frequency $\Omega_K$  estimated
at the inner disc radius, $R_I$.

The precession pulsation, $\omega_P$, is determined as follows:

\begin{equation}
 \omega_P= \frac{I_3}{I_1 cos\theta} \Omega_K,
\end{equation}




where

\begin{equation}
 I_1= I_2= \frac{1}{2} I_3=\Delta I
\end{equation}

\noindent and $I_3$=$I_D$. If $\theta$ is the wobble angle defined as the angle between the angular momentum and the symmetry 
axis of the disc, it is found that for  
$\theta$$\simeq$$0$, $cos$$\theta$$\simeq1$ and $\omega_P$$\simeq$2 $\Omega_K$=$3.01$$\times$$10^{-4}$ Hz.  

Therefore, the  precession period of the disc  detected by Spitzer would be: 

\begin{equation}
T_P =\frac{2 \pi}{\omega_P}=  20.9\ ks,
\end{equation}

\noindent in a  fair agreement with half period (27.5 ks) of the  X-ray modulation observed by NuSTAR from 4U 0142+61.

\begin{table*}
\caption{\label{table1} Gravitational wave amplitude generated by precessing discs and detection distances for  $\theta$=10$^o$. The signal-to-noise ratio (S/N) is inferred from the LISA (below 10$^{-3}$ Hz) and BBO (above 10$^{-3}$ Hz)  sensitivities reported in Figure 2. Maximum distances for precessing disc detection, when feasible, are set for S/N=1. The detection probability would increase with the square root of the observational time.}  
\label{tab:1}
\begin{tabular*}{13cm}{llllll}
\hline
 & Disc mass  &  $\omega_{1}$( $\omega_{2}$)  &   h$_{\omega_1}$(h$_{\omega_2}$)  & S/N$_{\omega_1}$(S/N$_{\omega_2}$)& Distance\\
  &  (kg)  &(Hz)   &   &   &(kpc)\\
\hline
 4U 0142+61 & 5.97$\times$10$^{25}$ &    3.1$\times$10$^{-4}$(6.2$\times$10$^{-4}$)    &  1.1$\times$10$^{-26}$(4.5$\times$10$^{-26}$)  & $<<$1 ($<<$1)  &  0.05(0.05)  \\
\hline
 Crab  & 5.97$\times$10$^{25}$  &   3.9$\times$10$^{-2}$(7.8$\times$10$^{-2}$) & 6.7$\times$10$^{-26}$(2.7$\times$10$^{-25}$)   & $<<$1 ($<$1)  &   0.05(0.05) \cr
       & 10$^{27}$              &  `` &   1.1$\times$10$^{-24}$(4.5$\times$10$^{-25}$) &  $<$1(1)  &  0.05(0.5)  \\
       & 10$^{28}$              &  `` &  9.3$\times$10$^{-25}$(4.8$\times$10$^{-25}$)&  1(1)&  0.6(4.65) \\
       & 10$^{29}$              &  `` &  8.7$\times$10$^{-25}$(4.5$\times$10$^{-25}$)&  1(1)&  6.8(50.0) \\
\hline
\end{tabular*}
\end{table*}

\begin{table*}
\caption{\label{table1} Same as Table 2 for $\theta$=30$^o$.   For  masses of 10$^{29}$ kg, in principle,  disc precession detection may extend to the whole Galaxy (WG).}
\label{tab:2}
\begin{tabular*}{13cm}{llllll}
\hline
  &  Disc mass  & $\omega_{1}$( $\omega_{2}$)  &  h$_{\omega_1}$(h$_{\omega_2}$)  & S/N$_{\omega_1}$(S/N$_{\omega_2}$)& Distance\\
   &  (kg)  &   (Hz)   &   &  &(kpc)\\
\hline
 4U 0142+61 & 5.97$\times$10$^{25}$ &    3.5$\times$10$^{-4}$(7.0$\times$10$^{-4}$)    &   1.9$\times$10$^{-25}$(4.8$\times$10$^{-25}$)& $<<$1 ($<<$1)  &   0.05(0.05)  \cr
\hline
 Crab  & 5.97$\times$10$^{25}$  &       4.4$\times$10$^{-2}$(8.8$\times$10$^{-2}$ ) & 7.2$\times$10$^{-25}$(4.8$\times$10$^{-25}$)   &   1(1)&    0.05(0.3)\cr
       & 10$^{27}$              &  `` & 8.0$\times$10$^{-25}$(4.2$\times$10$^{-25}$)&  1(1)&     0.75(5.7)\cr
       & 10$^{28}$              &  `` &  8.1$\times$10$^{-25}$(4.2$\times$10$^{-25}$)&  1(1)&   7.5(57.1)\cr
       & 10$^{29}$              &  `` &  7.1$\times$10$^{-25}$(4.1$\times$10$^{-25}$)&  1(1)&   84.5(WG)\cr
\hline
\end{tabular*}
\end{table*}



\section{Gravitational wave emission from  precessing discs}

In  \cite{gri16} and references therein it was shown that 
the frequencies
of gravitational waves generated by precessing circumpulsar discs  are $\omega_{GW}$ and 2$\omega_{GW}$, where $\omega_{GW}$ is defined  as follows:

\begin{equation}
 \omega_{GW}=  \omega_{P}.
\end{equation}

 
Under the small wobble angle approximation ($\theta$=10$^{\circ}$), the  gravitational wave frequencies associated with a Keplerian disc precessing around the magnetar 
 4U 0142+61 would be  3.1$\times$10$^{-4}$ Hz and 6.2$\times$10$^{-4}$ Hz that,  as it can be observed in Figure 2,  lie in the LISA sensitivity band, 
nominally  ranging between 10$^{-1}$ and 10$^{-4}$ Hz. It is worthwhile to point out that,  due to the LISA Pathfinder  \citep{armano18} 
encouraging results, the frequency range of the LISA observations may be likely extended down to 10$^{-5}$ Hz. 

The gravitational wave amplitudes  ($h$) generated by precessing discs depend on the inclination angle ($i$) of the disc angular momentum with respect to the line
of sight (see \cite{gri16} 
and references therein for details). For instance, in the case $i$$\simeq0$ and $\theta$ small 
\citep{lees}:

\begin{equation}
 h \simeq  \frac{G}{ c^4} \frac{\omega^2}{r} \Delta I  \theta^2,
\end{equation}

\noindent where {\it r}  is the distance between the precessing disc and the observer and $\Delta I$ is defined in equation 12.



The  maximum distance at which the future interferometers for gravitational wave detection in space will observe precessing discs depends on the principal moments of inertia which, in turn, depend on the disc masses and dimensions. 
 Mass discs up to 10$^{29}$ kg are considered here according to  \cite{qiao}  and references therein. The amplitudes of the gravitational waves and  associated frequencies for a precessing disc around  4U 0142+61  distant  3.6 kpc from the Solar System \citep{Olausen_2014} are reported in Table 1. In the same Table, similar estimates are reported for a disc possibly precessing around Crab  located at 2 kpc from Earth \citep{Manchester_2005} or a similar young pulsar, for which   different masses are considered. 
The inner and outer radii of the disc are set equal to 8$\times$10$^7$ m and 1.1$\times$10$^8$ m, respectively, as lower limits.
Wobble angles of 10 and 30 degrees are considered in Tables 1 and 2.  
 The S/N ratio reported in  the same tables appears $>$1 up to maximum distances of tens of kpc for disc masses larger than 10$^{28}$ kg and wobble angles of the order of tens of degrees. 
The maximum sensitivity of LISA is 2.5$\times$10$^{-22}$ at 3 mHz.
 It is found that  the disc around   4U 0142+61 cannot be detected by the LISA interferometer.  As it can be also noted in Figure 2.
ALIA is supposed to be characterized by a slightly better minimum sensitivity of 5.6$\times$ 10$^{-23}$ but in the frequency band 10$^{-2}$-10$^{-1}$ Hz.  DECIGO and BBO are designed for maximum sensitivities of  5.7$\times$10$^{-25}$ and  3.3$\times$10$^{-25}$, respectively, between 10$^{-2}$ Hz and  1 Hz. 
 In the case  of gravitational wave emission by precessing discs  with masses  up to 3.5 orders of magnitude larger than that formed around 4U 0142+61,  BBO or DECIGO would be sensitive enough for the detection  in case of wobble angles of a few tens of degrees. 
Moreover,  
the integration of  continuous gravitational wave   signals would improve the S/N ratio with the square root of time allowing us to resolve these sources with respect to intermediate mass black hole merging in the same frequency range.

\begin{figure}
\begin{center}
\includegraphics[width=\columnwidth]{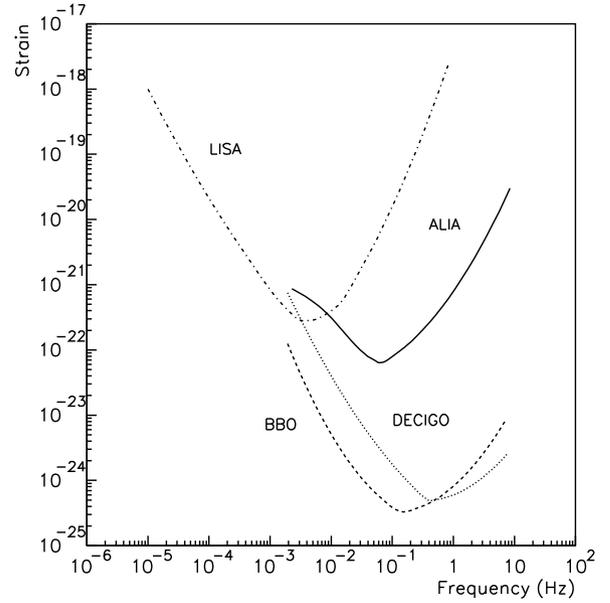}
\end{center}
\caption{\label{}  Comparison among  sensitivities of the future  interferometers for gravitational wave detection in space.}
\label{figure1}
\end{figure}

Even if the component of the  radiation reaction torque
 perpendicular to the angular momentum is taken into account, the wobble angle change can be expressed as a function of $\theta$ as follows:

\begin{equation}
\dot{\theta}= -\frac{1}{\tau_\theta} \theta,
\end{equation}

\noindent where:

\begin{equation}
\frac{1}{\tau_\theta} = \frac{2G}{5c^5} \frac{(\Delta I)^2} {I_1} \omega^4.
\end{equation}


The disc precession lifetime in all considered cases results to be larger than one million year. As a result, the actual capability of the space interferometers to detect precessing discs would depend on the space mission elapsed time.



\section{High-energy positron and electron propagation in the region of disc formation}

In order to estimate the effects of a  disc (precessing or not) formed around a young pulsar such as Crab in limiting the propagation of TeV electrons and positrons,  pulsar  and   disc blackbody radiations are considered and compared.

The blackbody photon density near a disc is estimated as follows:

\begin{equation}
 \frac{N}{V}=8 \pi \left(\frac{kT}{hc}\right)^3\times2.405 
\end{equation}
 
\noindent  
 for the temperatures of 920 K and 1800 K and V= 1 cm$^3$, N results to be 1.57$\times$10$^{10}$ photons cm$^{-3}$ and 1.18$\times$10$^{11}$ photons cm$^{-3}$, respectively.
The photon average energy associated with the disc temperature of 920 K and 1800 K is 0.21 eV and 0.43 eV in the near-infrared range on the basis of equation 2.
In the following we focus in particular on disc temperature of 1800 K to study the maximum effect of a disc in limiting  TeV energy electron and positron propagation. 


The pulsar blackbody radiation intensity estimated in the disc region according to the Stefan-Boltzmann law would be 2.7$\times$10$^7$  W m$^{-2}$. However, due to the higher energy of the radiation  with respect to that of the disc,  the photon density from the pulsar results to be  1.15$\times$10$^9$ photons cm$^{-3}$, two orders of magnitude smaller than that  near the disc. The infrared radiation density from the pulsar would be approximately one order of magnitude smaller than that of the disc \citep{penny}.
 
The Klein-Nishina IC total cross section  of TeV electrons and positrons scattering low-energy photons to TeV energies must be considered since the photon energy, h$\nu_o$, in the rest frame of electrons and positrons is  h$\nu_o$/(m$_e$c$^2$) $>$ 1 with m$_e$c$^2$ being the electron mass. It is worthwhile to recall that  the maximum photon energy in the electron rest frame is h$\nu_o$=h$\gamma$$\nu_L$(1+$\beta$) corresponding to a maximum frequency of 2$\gamma$$\nu_L$  where $\nu_L$ is the frequency of the photon in the laboratory frame, $\beta$=v/c  represents the electron speed with respect to the light speed and $\gamma$=$1/\sqrt{1-\beta^2}$. Since $\gamma$=5$\times$10$^6$ for the photons observed by MAGIC,  as a case study, the scattering  of infrared radiation leads to h$\nu_o$/(m$_e$c$^2$) $\simeq$$ 8 >$ 1.

 The total Klein-Nishina  cross section is given by:

\begin{equation}\begin{split}
\sigma_{KN}=&\sigma_T\frac{3}{4} \left[\frac{1+x} {x^3} \left( \frac{2x(1+x)} {1+2x}-ln(1+2x)\right)+ \frac{1}{2x} ln(1+2x)+ \right.\\ 
&\left. -\frac{1+3x}{(1+2x)^2} \right] 
\end{split}\end{equation}

\noindent where $x$= h$\nu_o$/(m$_e$ c$^2$). Since  h$\nu_o$/(m$_e$ c$^2$)$>$1 the
above equation can be reduced as follows:

\begin{equation}
\sigma_{KN}=\sigma_T \frac{3}{8} \frac{1}{x} \left(ln2x + \frac{1}{2} \right).
\end{equation}    


The total cross section and associated collision length would be 9.9$\times$10$^{-30}$ m$^2$ and 8.6$\times$$10^{11}$ m, respectively.
Only discs of AU dimensions similar to that of RX J0806.4-4123 with temperatures of at least hundreds of degrees  would contribute to TeV e$^+$ and e$^-$  IC scattering of infrared photons to TeV energies. 
 No claim can be made of a large disc surrounding Crab due to continuous electron synchrotron emission observed to dominate the infrared emission in the range 3.6-4.5 $\mu$m  while 24-70 $\mu$m emission is associated with less than one solar mass dust  in the Nebula  \citep{temim_2006,lyne2014}. 

\section{Conclusions}

In the next few years, the James Webb telescope will open a new window on the infrared Universe, by shedding light on supernova fallback disc formation around pulsars. Simulations indicate that  fallback disc formation may likely occur in a large sample of pulsars and magnetars. 
A modulated hard X-ray emission from the AXP 4U 0142+61 observed by the NuSTAR experiment is consistent with a precessing  Keplerian disc formed  around the magnetar. 

Gravitational waves generated by the precession of a disc similar to that observed around 4U 0142+61, would be characterized by frequencies belonging to the LISA detection band but also by amplitudes lying below the sensitivity of the interferometer, while gravitational waves generated by precessing discs  formed around young pulsars appear detectable with future generation BBO and DECIGO space interferometers up to tens of kpc  distance and beyond in case of disc large masses, hundreds of kelvin temperature and
wobble angle of tens of degrees. It was also shown that the infrared radiation emission from discs of AU dimensions may contribute to IC scattering  affecting high-energy electron and positron propagation in the pulsar near environment.

\section*{Acknowledgements}
This research work was funded by the Department of Pure and Applied Sciences of the University of Urbino Carlo Bo.
The author thanks M. Fabi of the University of Urbino Carlo Bo for technical support and Dr. F. Sabbatini for reading the manuscript.

\section*{Data Availability}
No new data were generated or analysed in support of this research.
 



\bibliographystyle{mnras}
\bibliography{magnetar_mnras_final_u} 

\begin{thebibliography}{}
\makeatletter
\relax
\def\mn@urlcharsother{\let\do\@makeother \do\$\do\&\do\#\do\^\do\_\do\%\do\~}
\def\mn@doi{\begingroup\mn@urlcharsother \@ifnextchar [ {\mn@doi@}
  {\mn@doi@[]}}
\def\mn@doi@[#1]#2{\def\@tempa{#1}\ifx\@tempa\@empty \href
  {http://dx.doi.org/#2} {doi:#2}\else \href {http://dx.doi.org/#2} {#1}\fi
  \endgroup}
\def\mn@eprint#1#2{\mn@eprint@#1:#2::\@nil}
\def\mn@eprint@arXiv#1{\href {http://arxiv.org/abs/#1} {{\tt arXiv:#1}}}
\def\mn@eprint@dblp#1{\href {http://dblp.uni-trier.de/rec/bibtex/#1.xml}
  {dblp:#1}}
\def\mn@eprint@#1:#2:#3:#4\@nil{\def\@tempa {#1}\def\@tempb {#2}\def\@tempc
  {#3}\ifx \@tempc \@empty \let \@tempc \@tempb \let \@tempb \@tempa \fi \ifx
  \@tempb \@empty \def\@tempb {arXiv}\fi \@ifundefined
  {mn@eprint@\@tempb}{\@tempb:\@tempc}{\expandafter \expandafter \csname
  mn@eprint@\@tempb\endcsname \expandafter{\@tempc}}}

\bibitem[\protect\citeauthoryear{Aguilar et~al.,}{Aguilar
  et~al.}{2019}]{amspos}
Aguilar M.,  et~al., 2019, \mn@doi [Phys. Rev. Lett.]
  {10.1103/PhysRevLett.122.041102}, 122, 041102

\bibitem[\protect\citeauthoryear{Aguilera, Pons  \& Miralles}{Aguilera
  et~al.}{2008}]{aguilera}
Aguilera D.~N.,  Pons J.~A.,   Miralles J.~A.,  2008, \mn@doi [The
  Astrophysical Journal] {10.1086/527547}, 673, L167

\bibitem[\protect\citeauthoryear{Alpar}{Alpar}{2001}]{Alpar_2001}
Alpar M.~A.,  2001, \mn@doi [The Astrophysical Journal] {10.1086/321393}, 554,
  1245

\bibitem[\protect\citeauthoryear{{Amaro-Seoane} et~al.,}{{Amaro-Seoane}
  et~al.}{2017}]{lisa}
{Amaro-Seoane} P.,  et~al., 2017, arXiv e-prints, \href
  {https://ui.adsabs.harvard.edu/abs/2017arXiv170200786A} {p. arXiv:1702.00786}

\bibitem[\protect\citeauthoryear{{Ansoldi} et~al.,}{{Ansoldi}
  et~al.}{2016}]{ansoldi}
{Ansoldi} S.,  et~al., 2016, \mn@doi [\aap] {10.1051/0004-6361/201526853},
  \href {https://ui.adsabs.harvard.edu/abs/2016A&A...585A.133A} {585, A133}

\bibitem[\protect\citeauthoryear{{Armano} et~al.,}{{Armano}
  et~al.}{2018}]{armano18}
{Armano} M.,  et~al., 2018, \mn@doi [Physical Review Letters]
  {10.1103/PhysRevLett.120.061101}, \href
  {http://adsabs.harvard.edu/abs/2018PhRvL.120f1101A} {120, 061101}

\bibitem[\protect\citeauthoryear{{Attallah}}{{Attallah}}{2016}]{attallah}
{Attallah} R.,  2016, \mn@doi [\jcap] {10.1088/1475-7516/2016/12/025}, \href
  {https://ui.adsabs.harvard.edu/abs/2016JCAP...12..025A} {2016, 025}

\bibitem[\protect\citeauthoryear{Benli \& Ertan}{Benli \&
  Ertan}{2017}]{Benli_2017}
Benli O.,  Ertan {\" U}.,  2017, \mn@doi [Monthly Notices of the Royal
  Astronomical Society] {10.1093/mnras/stx1735}, 471, 2553

\bibitem[\protect\citeauthoryear{Benli \& Ertan}{Benli \&
  Ertan}{2018}]{Benli_2018}
Benli O.,  Ertan {\" U}.,  2018, \mn@doi [Monthly Notices of the Royal
  Astronomical Society] {10.1093/mnras/sty1399}, 478, 4890

\bibitem[\protect\citeauthoryear{Benli, {\c{C}}ali{\c{s}}kan  \& Ertan}{Benli
  et~al.}{2015}]{beinli}
Benli O.,  {\c{C}}ali{\c{s}}kan {\c{S}}.,   Ertan {\" U}.,  2015, \mn@doi
  [Monthly Notices of the Royal Astronomical Society] {10.1093/mnras/stu2569},
  447, 2282

\bibitem[\protect\citeauthoryear{Chatterjee, Hernquist  \& Narayan}{Chatterjee
  et~al.}{2000}]{chatterjee}
Chatterjee P.,  Hernquist L.,   Narayan R.,  2000, \mn@doi [The Astrophysical
  Journal] {10.1086/308748}, 534, 373

\bibitem[\protect\citeauthoryear{Currie \& Hansen}{Currie \&
  Hansen}{2007}]{currie}
Currie T.,  Hansen B.,  2007, \mn@doi [The Astrophysical Journal]
  {10.1086/520327}, 666, 1232

\bibitem[\protect\citeauthoryear{Cutler \& Harms}{Cutler \&
  Harms}{2006}]{cutler}
Cutler C.,  Harms J.,  2006, \mn@doi [Phys. Rev. D]
  {10.1103/PhysRevD.73.042001}, 73, 042001

\bibitem[\protect\citeauthoryear{Diesing \& Caprioli}{Diesing \&
  Caprioli}{2020}]{caprioli}
Diesing R.,  Caprioli D.,  2020, \mn@doi [Phys. Rev. D]
  {10.1103/PhysRevD.101.103030}, 101, 103030

\bibitem[\protect\citeauthoryear{Durant \& van Kerkwijk}{Durant \& van
  Kerkwijk}{2006}]{durant}
Durant M.,  van Kerkwijk M.~H.,  2006, \mn@doi [The Astrophysical Journal]
  {10.1086/507605}, 652, 576

\bibitem[\protect\citeauthoryear{Erdeve, Kalemci  \& Alpar}{Erdeve
  et~al.}{2009}]{erdeve}
Erdeve I.,  Kalemci E.,   Alpar M.~A.,  2009, \mn@doi [The Astrophysical
  Journal] {10.1088/0004-637x/696/2/1792}, 696, 1792

\bibitem[\protect\citeauthoryear{Ertan, Erkut, Ek{\c{s}}i  \& Alpar}{Ertan
  et~al.}{2007}]{Ertan_2007}
Ertan {\" U}.,  Erkut M.~H.,  Ek{\c{s}}i K.~Y.,   Alpar M.~A.,  2007, \mn@doi
  [The Astrophysical Journal] {10.1086/510303}, 657, 441

\bibitem[\protect\citeauthoryear{Ertan, Ek{\c{s}}i, Erkut  \& Alpar}{Ertan
  et~al.}{2009}]{Ertan_2009}
Ertan {\" U}.,  Ek{\c{s}}i K.~Y.,  Erkut M.~H.,   Alpar M.~A.,  2009, \mn@doi
  [The Astrophysical Journal] {10.1088/0004-637x/702/2/1309}, 702, 1309

\bibitem[\protect\citeauthoryear{Ertan, {\c{C}}ali{\c{s}}kan, Benli  \&
  Alpar}{Ertan et~al.}{2014}]{Ertan_2014}
Ertan {\" U}.,  {\c{C}}ali{\c{s}}kan {\c{S}}.,  Benli O.,   Alpar M.~A.,  2014,
  \mn@doi [Monthly Notices of the Royal Astronomical Society]
  {10.1093/mnras/stu1523}, 444, 1559

\bibitem[\protect\citeauthoryear{Feng \& Zhang}{Feng \& Zhang}{2018}]{feng}
Feng J.,  Zhang H.-H.,  2018, \mn@doi [The Astrophysical Journal]
  {10.3847/1538-4357/aabf87}, 858, 116

\bibitem[\protect\citeauthoryear{{Gardner} et~al.,}{{Gardner}
  et~al.}{2006}]{gardner}
{Gardner} J.~P.,  et~al., 2006, \mn@doi [\ssr] {10.1007/s11214-006-8315-7},
  \href {https://ui.adsabs.harvard.edu/abs/2006SSRv..123..485G} {123, 485}

\bibitem[\protect\citeauthoryear{Gen{\c{c}}ali \& Ertan}{Gen{\c{c}}ali \&
  Ertan}{2020}]{Gencali_2020}
Gen{\c{c}}ali A.~A.,  Ertan {\" U}.,  2020, \mn@doi [Monthly Notices of the
  Royal Astronomical Society] {10.1093/mnras/staa3371}, 500, 3281

\bibitem[\protect\citeauthoryear{{Grimani}}{{Grimani}}{2004}]{gri04}
{Grimani} C.,  2004, \mn@doi [\aap] {10.1051/0004-6361:20040044}, \href
  {https://ui.adsabs.harvard.edu/abs/2004A&A...418..649G} {418, 649}

\bibitem[\protect\citeauthoryear{{Grimani}}{{Grimani}}{2007}]{gri07}
{Grimani} C.,  2007, \mn@doi [\aap] {10.1051/0004-6361:20077776}, \href
  {https://ui.adsabs.harvard.edu/abs/2007A&A...474..339G} {474, 339}

\bibitem[\protect\citeauthoryear{Grimani}{Grimani}{2013}]{gri13}
Grimani C.,  2013, \mn@doi [Advances in Space Research]
  {https://doi.org/10.1016/j.asr.2011.03.006}, 51, 322

\bibitem[\protect\citeauthoryear{Grimani}{Grimani}{2016}]{gri16}
Grimani C.,  2016, \mn@doi [Monthly Notices of the Royal Astronomical Society]
  {10.1093/mnras/stw1057}, 460, 2186

\bibitem[\protect\citeauthoryear{G\"{u}ver, \"{O}zel  \&
  G\"{o}{\u{g}}\"{u}{\c{s}}}{G\"{u}ver et~al.}{2008}]{guvert}
G\"{u}ver T.,  \"{O}zel F.,   G\"{o}{\u{g}}\"{u}{\c{s}} E.,  2008, \mn@doi [The
  Astrophysical Journal] {10.1086/525840}, 675, 1499

\bibitem[\protect\citeauthoryear{{Harding} \& {Ramaty}}{{Harding} \&
  {Ramaty}}{1987}]{harding}
{Harding} A.~K.,  {Ramaty} R.,  1987, in International Cosmic Ray Conference.
  p.~92

\bibitem[\protect\citeauthoryear{Hawley \& Krolik}{Hawley \&
  Krolik}{2018}]{refh}
Hawley J.~F.,  Krolik J.~H.,  2018, \mn@doi [The Astrophysical Journal]
  {10.3847/1538-4357/aadf90}, 866, 5

\bibitem[\protect\citeauthoryear{Huang, Liu, Joshi  \& Wang}{Huang
  et~al.}{2020}]{huang}
Huang Z.-Q.,  Liu R.-Y.,  Joshi J.~C.,   Wang X.-Y.,  2020, \mn@doi [The
  Astrophysical Journal] {10.3847/1538-4357/ab88cb}, 895, 53

\bibitem[\protect\citeauthoryear{Jiang \& Li}{Jiang \& Li}{2005}]{jiang}
Jiang Z.-B.,  Li X.-D.,  2005, ] {10.1088/1009-9271/5/5/006}, 5, 487

\bibitem[\protect\citeauthoryear{Kaplan, Chakrabarty, Wang  \& Wachter}{Kaplan
  et~al.}{2009}]{kaplan}
Kaplan D.~L.,  Chakrabarty D.,  Wang Z.,   Wachter S.,  2009, \mn@doi [The
  Astrophysical Journal] {10.1088/0004-637x/700/1/149}, 700, 149

\bibitem[\protect\citeauthoryear{{Kawamura} et~al.,}{{Kawamura}
  et~al.}{2020}]{decigo}
{Kawamura} S.,  et~al., 2020, arXiv e-prints, \href
  {https://ui.adsabs.harvard.edu/abs/2020arXiv200613545K} {p. arXiv:2006.13545}

\bibitem[\protect\citeauthoryear{Lee, Lee  \& Kim}{Lee et~al.}{2004}]{lees}
Lee H.~K.,  Lee C.~H.,   Kim H.~S.,  2004, Journal of the Korean Physical
  Society, 45, 564

\bibitem[\protect\citeauthoryear{Li}{Li}{2006}]{li06}
Li X.-D.,  2006, \mn@doi [Astrophys. J.] {10.1086/506962}, 646, L139

\bibitem[\protect\citeauthoryear{Lu \& Cheng}{Lu \& Cheng}{2002}]{lu}
Lu Y.,  Cheng K.~S.,  2002, \mn@doi [Chinese Journal of Astronomy and
  Astrophysics] {10.1088/1009-9271/2/2/161}, 2, 161

\bibitem[\protect\citeauthoryear{Lyne, Jordan, Graham-Smith, Espinoza, Stappers
   \& Weltevrede}{Lyne et~al.}{2014}]{lyne2014}
Lyne A.~G.,  Jordan C.~A.,  Graham-Smith F.,  Espinoza C.~M.,  Stappers B.~W.,
   Weltevrede P.,  2014, \mn@doi [Monthly Notices of the Royal Astronomical
  Society] {10.1093/mnras/stu2118}, 446, 857

\bibitem[\protect\citeauthoryear{Makishima, Murakami, Enoto  \&
  Nakazawa}{Makishima et~al.}{2018}]{makishima}
Makishima K.,  Murakami H.,  Enoto T.,   Nakazawa K.,  2018, \mn@doi
  [Publications of the Astronomical Society of Japan] {10.1093/pasj/psy129}, 71

\bibitem[\protect\citeauthoryear{Manchester, Hobbs, Teoh  \& Hobbs}{Manchester
  et~al.}{2005}]{Manchester_2005}
Manchester R.~N.,  Hobbs G.~B.,  Teoh A.,   Hobbs M.,  2005, \mn@doi [The
  Astronomical Journal] {10.1086/428488}, 129, 1993

\bibitem[\protect\citeauthoryear{Manconi, Di~Mauro  \& Donato}{Manconi
  et~al.}{2020}]{manconi}
Manconi S.,  Di~Mauro M.,   Donato F.,  2020, \mn@doi [Phys. Rev. D]
  {10.1103/PhysRevD.102.023015}, 102, 023015

\bibitem[\protect\citeauthoryear{Menou, Perna  \& Hernquist}{Menou
  et~al.}{2001}]{menou}
Menou K.,  Perna R.,   Hernquist L.,  2001, \mn@doi [The Astrophysical Journal]
  {10.1086/320927}, 554, L63

\bibitem[\protect\citeauthoryear{Ni}{Ni}{2016}]{wtni}
Ni W.-T.,  2016, \mn@doi [International Journal of Modern Physics D]
  {10.1142/S0218271816030012}, 25, 1603001

\bibitem[\protect\citeauthoryear{Olausen \& Kaspi}{Olausen \&
  Kaspi}{2014}]{Olausen_2014}
Olausen S.~A.,  Kaspi V.~M.,  2014, \mn@doi [The Astrophysical Journal
  Supplement Series] {10.1088/0067-0049/212/1/6}, 212, 6

\bibitem[\protect\citeauthoryear{{\" O}zcan, Gen{\c c}ali  \& Ertan}{{\" O}zcan
  et~al.}{2020}]{xdins}
{\" O}zcan {\c S}.,  Gen{\c c}ali A.~A.,   Ertan {\" U}.,  2020, \mn@doi
  [Monthly Notices of the Royal Astronomical Society] {10.1093/mnras/staa2493},
  498, 674

\bibitem[\protect\citeauthoryear{Penny}{Penny}{1982}]{penny}
Penny A.~J.,  1982, \mn@doi [Monthly Notices of the Royal Astronomical Society]
  {10.1093/mnras/198.3.773}, 198, 773

\bibitem[\protect\citeauthoryear{Perna, Hernquist  \& Narayan}{Perna
  et~al.}{2000}]{perna}
Perna R.,  Hernquist L.,   Narayan R.,  2000, \mn@doi [The Astrophysical
  Journal] {10.1086/309404}, 541, 344

\bibitem[\protect\citeauthoryear{Posselt, Pavlov, Ertan, {\c{C}}ali{\c{s}}kan,
  Luhman  \& Williams}{Posselt et~al.}{2018}]{posselt}
Posselt B.,  Pavlov G.~G.,  Ertan {\"U}.,  {\c{C}}ali{\c{s}}kan {\c{S}}.,
  Luhman K.~L.,   Williams C.~C.,  2018, \mn@doi [The Astrophysical Journal]
  {10.3847/1538-4357/aad6df}, 865, 1

\bibitem[\protect\citeauthoryear{{Qiao}, {Xue, Y. Q.}, {Xu, R. X.}, {Wang, H.
  G.}  \& {Xiao, B. W.}}{{Qiao} et~al.}{2003}]{qiao}
{Qiao} {Xue, Y. Q.} {Xu, R. X.} {Wang, H. G.}  {Xiao, B. W.} 2003, \mn@doi
  [A\&A] {10.1051/0004-6361:20031055}, 407, L25

\bibitem[\protect\citeauthoryear{{Shakura} \& {Sunyaev}}{{Shakura} \&
  {Sunyaev}}{1973}]{shakura}
{Shakura} N.~I.,  {Sunyaev} R.~A.,  1973, \aap, \href
  {https://ui.adsabs.harvard.edu/abs/1973A&A....24..337S} {500, 33}

\bibitem[\protect\citeauthoryear{Steiner, Heinke, Bogdanov, Li, Ho, Bahramian
  \& Han}{Steiner et~al.}{2018}]{steiner}
Steiner A.~W.,  Heinke C.~O.,  Bogdanov S.,  Li C.~K.,  Ho W. C.~G.,  Bahramian
  A.,   Han S.,  2018, \mn@doi [Monthly Notices of the Royal Astronomical
  Society] {10.1093/mnras/sty215}, 476, 421

\bibitem[\protect\citeauthoryear{Temim et~al.,}{Temim
  et~al.}{2006}]{temim_2006}
Temim T.,  et~al., 2006, \mn@doi [The Astronomical Journal] {10.1086/507076},
  132, 1610

\bibitem[\protect\citeauthoryear{Tennant et~al.,}{Tennant
  et~al.}{2001}]{tennant}
Tennant A.~F.,  et~al., 2001, \mn@doi [The Astrophysical Journal]
  {10.1086/321718}, 554, L173

\bibitem[\protect\citeauthoryear{{Tong}, {Song}  \& {Xu}}{{Tong}
  et~al.}{2011}]{tong}
{Tong} H.,  {Song} L.~M.,   {Xu} R.~X.,  2011, \mn@doi [\apj]
  {10.1088/0004-637X/738/1/31}, \href
  {http://cdsads.u-strasbg.fr/abs/2011ApJ...738...31T} {738, 31}

\bibitem[\protect\citeauthoryear{{Tong}, {Wang}  \& {Wang}}{{Tong}
  et~al.}{2020}]{tong1}
{Tong} H.,  {Wang} W.,   {Wang} H.-G.,  2020, \mn@doi [Research in Astronomy
  and Astrophysics] {10.1088/1674-4527/20/9/142}, \href
  {https://ui.adsabs.harvard.edu/abs/2020RAA....20..142T} {20, 142}

\bibitem[\protect\citeauthoryear{{Wang}, {Chakrabarty}  \& {Kaplan}}{{Wang}
  et~al.}{2006}]{wang}
{Wang} Z.,  {Chakrabarty} D.,   {Kaplan} D.~L.,  2006, \mn@doi [\nat]
  {10.1038/nature04669}, \href
  {http://adsabs.harvard.edu/abs/2006Natur.440..772W} {440, 772}

\bibitem[\protect\citeauthoryear{{Wolszczan}}{{Wolszczan}}{2008}]{wolszczan}
{Wolszczan} A.,  2008, \mn@doi [Physica Scripta Volume T]
  {10.1088/0031-8949/2008/T130/014005}, \href
  {https://ui.adsabs.harvard.edu/abs/2008PhST..130a4005W} {130, 014005}

\bibitem[\protect\citeauthoryear{{Zhang} \& {Cheng}}{{Zhang} \&
  {Cheng}}{2001}]{zhang}
{Zhang} L.,  {Cheng} K.~S.,  2001, \mn@doi [\aap] {10.1051/0004-6361:20010021},
  \href {https://ui.adsabs.harvard.edu/abs/2001A&A...368.1063Z} {368, 1063}

\makeatother
\end{thebibliography}








\bsp	
\label{lastpage}
\end{document}